**Simplifying concentration-polarization of trace-ions in pressure-driven membrane processes**

Yaeli S. Oren[1], Viatcheslav Freger[2], Oded Nir[*,1]

[1] Department of Desalination and Water Treatment, Zuckerberg Institute for Water Research,

The Jacob Blaustein Institutes for Desert Research, Ben-Gurion University of the Negev,
Sede-Boqer Campus 8499000, Israel

[2] Wolfson Department of Chemical Engineering, Technion – IIT, Haifa 32000, Israel

Corresponding Author Email: odni@bgu.ac.il (O.N.)

## Abstract

Accounting for concentration-polarization (CP) is critical for modeling solute transport in membrane separation processes. In a mixed-electrolyte solution, ions CP is affected not only by diffusion and advection but also by electromigration. Yet, the classic film model, lacking an electromigration term, is frequently used for modeling ion CP. Often, ion CP is altogether neglected to reduce the computational load. Here, we study the CP of trace ions in a dominant salt solution, a case relevant for many reverse-osmosis and nanofiltration processes. First, we revisit the solution-diffusion-electromigration-film theory to obtain an analytical solution for the CP and membrane-transport of trace-ions in a dominant salt solution. Secondly, we consider limiting conditions relevant to reverse-osmosis and nanofiltration, from which we derive two compact equations that emerge as a seamless extension to the classic film theory. These equations can be used to account for the effect of electromigration on CP with minimal effort. Thirdly, we use our theory to quantify the effect of electromigration on ion CP in different dominant salt solutions. Finally, by analyzing two environmental membrane processes, we demonstrate how our theory deviates from the conventional one and quantify the implications on membrane scaling potential and the transport of ionic contaminants.

## 1. Introduction

Concentration polarization (CP) is inherent in pressure-driven membrane processes and may critically affect performance parameters such as permeate flux, solute rejection, and membrane fouling. For salt rejecting membranes, CP is mainly described by the conventional film theory, which balances advective and diffusive solute fluxes at the unstirred layer adjacent to the membrane surface. While the film theory is an accurate physical description for CP of uncharged solutes and single salt solutions [1], it is not valid for ions in practically relevant mixed electrolyte solutions (comprising three ions or more). This discrepancy emanates from the extended Nernst-Planck theory, which describes ion transport through the sum of advective, diffusive, and *electromigrative* fluxes. Since electromigration depends on the ions' fluxes in the solution, the general description of ion transport in the CP layer is an inter-dependent system of partial differential equations. A straightforward approach is to solve this system of equations numerically (coupled to the membrane transport equations) [1–6]. However, due to the complexity of the problem, in many previous membrane process simulation studies – focusing on either trace contaminants removal or chemical fouling – electromigration in the CP layer is not considered [7–9]. Instead, the problem is oversimplified by either invoking the conventional film model, considering only a single salt, or neglecting CP altogether [10–21]. Here, we show that electromigration can significantly affect the CP of ions in a mixed-electrolyte solution and for practically relevant conditions.

In many practical applications of nanofiltration (NF) and reverse osmosis (RO) membranes, the treated water can be adequately described as a solution containing one dominant salt and one or more trace-level ions. In such ion mixtures, the electric field that drives the electromigration of trace ions can be treated as a function of the dominant ions alone, as shown in Figure 1a. This approximation simplifies the mathematical problem by breaking much of the interdependency between different ion fluxes, allowing for the derivation of a general analytical solution. Indeed, in 2011 Yaroshchuk *et al*. [22] proposed an electromigration-film-based CP description for such salt mixtures. However, it was scarcely used [6,23,24], probably due to the following (1) CP was not the focus of that work; (2) a complete analytical solution was not given, and the mathematical description was quite cumbersome. Here, we focused only on the effect of electromigration in the boundary layer on trace-ions CP. Starting from the fundamental transport equations, we developed a complete analytical solution. We also derived new compact CP equations for two limiting cases that are practically relevant for many RO and NF applications. These simple equations can be used directly to estimate CP or implemented in process simulations. We used these equations to explore how electromigration affects the CP of trace ions and highlight two cases where neglecting electromigration could hamper the estimation of both scaling tendency and trace-ion passage.

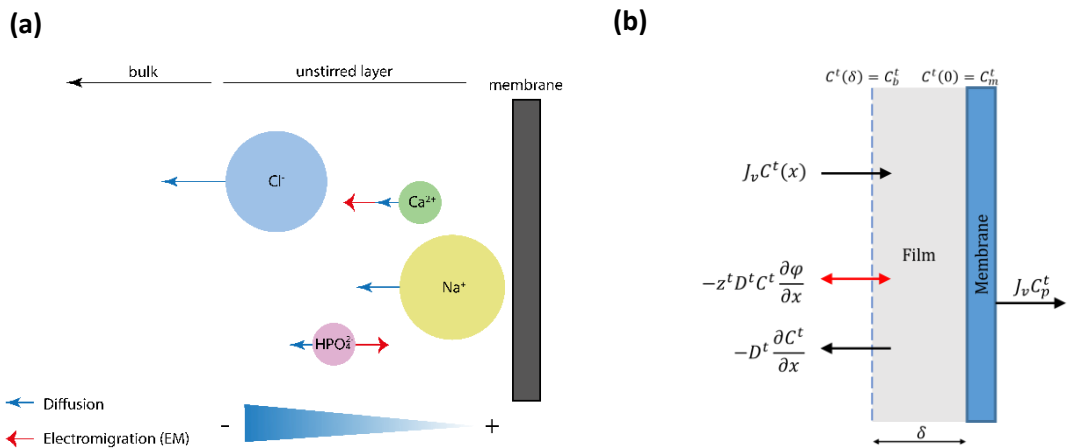

**Figure 1.** (a) Trace ion ($Ca^{2+}$, $HPO_4^{2-}$) migration induced by an electric field formed by a dominant salt (*NaCl*). Due to the slower back-diffusion of sodium, an induced electrical field appears, driving trace anion towards the membrane and trace cation away from the membrane. (b) Schematic description of trace ion mass transport in the unstirred layer of thickness δ.

## 2. Theory

We start from the extended Nernst-Plank transport equation describing trace ion(s) flux in the unstirred layer (Figure 1b),

$$J_v C^t(x) = -D_t \frac{\partial C^t(x)}{\partial x} - D_t z_t C^t(x) \frac{\partial \varphi(x)}{\partial x} + J_v C_p^t \quad (1)$$

where $J_v$ is the volumetric flux, $x$ is the flow direction perpendicular to the membrane, C is concentration, $D$ is the diffusion coefficient, $z$ is the charge number, and $\varphi$ is the dimensionless electric potential. The subscription '$p$' stands for permeate and the superscript '$t$' for trace ion. Considering also electroneutrality, which, by definition, is maintained by the dominant salt, the following analytical solution for trace ion CP can be derived:

$$\frac{C_m^t}{C_b^t} = e^{\frac{J_v \delta}{D_t}} \left( R_s e^{\frac{J_v \delta}{D_s}} + 1 - R_s \right)^{z_t \theta_\delta} - \frac{C_p^t}{C_b^t} \left( 1 + \frac{R_s e^{\frac{J_v \delta}{D_s}}}{1 - R_s} \right)^{z_t \theta_\delta} \left( e^{\frac{J_v \delta}{D_t}} F_1 - F_2 \right), \quad (2)$$

where,

$$F_1 = {}_2F_1\left(z_t \theta_\delta, -\frac{D_s}{D_t}; 1 - \frac{D_s}{D_t}; \frac{-R_s}{1-R_s}\right), \quad (3a)$$

$$F_2 = {}_2F_1\left(z_t \theta_\delta, -\frac{D_s}{D_t}; 1 - \frac{D_s}{D_t}; \frac{-R_s e^{\frac{J_v \delta}{D_s}}}{1-R_s}\right). \quad (3b)$$

$$\theta_\delta = \frac{D_s^+ - D_s^-}{z_+ D_s^+ - z_- D_s^-}, \quad (4)$$

$$D_s = \frac{(z_+ - z_-) D_s^+ D_s^-}{z_+ D_s^+ - z_- D_s^-} \quad (5)$$

and the dominant salt rejection $R_s$ is given by:

$$R_s = 1 - \frac{C_p^s}{C_b^s}. \tag{6}$$

The superscripts '$s$', '$+$' and '$-$' represent the dominant cation and anion, respectively. Complete derivation and further explanations are in the SI. The solution includes the Gauss hypergeometric function $_2F_1$ - an infinite series sum that could be estimated using existing numerical algorithms (Python code attached in the SI).

The compact membrane transport equation derived in our previous work [25] completes the mathematical description of cross-membrane trace ion transport in a dominant salt solution within the solution-diffusion-electromigration-film (SDEF) framework.

$$C_p^t = \frac{C_m^t \omega_t}{\omega_t (1-R_s)^{-z_t \theta_m} + J_v \frac{1 - (1-R_s)^{1-z_t \theta_m}}{R_s(1-z_t \theta_m)}}. \tag{7}$$

Eq. 2 (where $\omega_t$ is the trace-ion permeability and Eq. 7) form the complete general analytical solution for NF and RO of trace ions in dominant salt solutions.

From Eq. 2, we can derive two compact and practical limiting cases. First, we look at a case where the rejection of both the dominant and trace ions approach unity – a good approximation for RO. Eq. 2 then reduces to a multiplication of two exponents, and we arrive at the following remarkably simple compact expression for trace ions CP.

$$\frac{C_m^t}{C_b^t} = e^{\frac{J_v \delta}{D_t}} \boldsymbol{e^{\frac{J_v \delta z_t \theta_\delta}{D_s}}}. \tag{8}$$

In this new useful approximate expression, the first exponent coincides with the conventional film theory for neutral solutes and single salts (with complete rejection). The second exponent (highlighted in bold) conveniently corrects for the effect of electromigration in trace-ions CP. This electromigration correction factor can either increase or decrease the extent of CP, depending on the signs of $z_t$ and $\theta_\delta$ (both can be either positive or negative).

In the second limiting case, we consider a complete rejection of trace-ions with a partial rejection of the dominant salt. Such a scenario is a good approximation in many NF applications, where large or multivalent ions are separated from a monovalent dominant salt solution, e.g., removing trace ionic contaminants or scaling ions [26–30]. For such cases, Eq. 2 reduces to

$$\frac{C_m^t}{C_b^t} = e^{\frac{J_v\delta}{D_t}} \left( R_s e^{\frac{J_v\delta}{D_s}} + 1 - R_s \right)^{z_t\theta_\delta}. \qquad (9)$$

Here, the electromigration concentration factor expands to include the effect of partial rejection of the dominant salt ($R_S$). These simple yet theoretically established correction factors can be easily implemented for quantifying CP of trace ions, thus increasing the likelihood of adoption in both research and industry.

## 3. Methods

### 3.1 Calculations and numerical implementation

To calculate the concentration-polarization of trace-ions ($Ca^{2+}$, $HPO_4^{2-}$ and $HAsO_4^-$) in different dominant salts (Fig. 2), we used the analytical solution for trace ion CP (Eq. 2). To estimate the effect of electromigration, we compared Eq. 2 to the conventional film model. The Gauss hypergeometric functions ($_2F_1$) appearing in Eq. 2 were solved using the Python language SciPy library's [31] built-in function. Similarly, we used Eq. 2 to calculate CP of trace ions in practical membrane applications (Fig. 3) and compared the results to the conventional film theory and the relevant limiting cases (Eq. 8 for RO and Eq. 9 for NF). We used a PHREEQC-Python coupled code to calculate the precipitation potential of amorphous calcium phosphate (in case 1) and calcium sulfate (in case 2) in the modeled concentration-polarization layer. The dominant salt CP was calculated using the conventional film theory. Python 2.7 was coupled to PHREEQC 3.5 using a freely available module. Minteq.v4 and Wateq4f thermochemical databases were used for case study 1 (Fig. 3a&b) and case study 2 (Fig. 3c&d), respectively. More details on the calculations and all the Python and PHREEQC scripts are in the SI.

### 3.2 Description of the general case studies

The practical implications of trace-ions electromigration in the boundary layer were evaluated in terms of chemical scaling propensity and solute passage by resolving CP in the flux domain and comparing the full analytical solution (Eq. 2 and Eq. 7) to both the limiting cases (Eq. 8 and Eq. 9) and the conventional film model (Eq. S13). The parameters needed for calculating the permeation of trace ions ($\omega_i$, $\theta_m$) were constant for each membrane, thus allowing a valid comparison between the different CP approximations. Two general case studies were chosen to examine the impact of electromigration on CP modeling. In the first case study, we analyzed a final stage of RO treating wastewater effluent, reaching a recovery ratio of ~90%. The dominant salt was sodium chloride (85 mM), and the examined trace ions were calcium (10 mM), monohydrogen phosphate (0.5 mM $HPO_4^{2-}$, dominant at typical wastewater pH values of 7-8), and ammonium (15 mM). Calcium and phosphate may precipitate as ACP (amorphous calcium phosphate), which is the least kinetically inhibited Ca-P solid phase under these conditions [32]. Ammonium is a relatively mobile ion that could, undesirably, permeate through the membrane. In the second case study, we examined arsenate removal from simulated brackish groundwater by two types of NF membranes, one tight (NF90) and one more loose (NF270). The feed solution, in this case, contained sodium chloride (170 mM) as the dominant salt and arsenic (0.001 mM) as trace, in accordance to Boussouga *et al*. [19] Feed solution composition and model parameters for both case studies appear in Tables S1 and S2.

## 4   Results and Discussion

Applying the conventional film theory to trace ions ignores the electromigrative flux in the boundary layer, which may result in either overestimation or underestimation of CP. To quantify this discrepancy, we calculated the CP modulus for trace divalent anions ($HPO_4^{2-}$) and cations ($Ca^{2+}$) in four different dominant salt solutions undergoing RO treatment (Fig. 2). The boundary layer thickness was in the range $10^{-5} < \delta < 1.5 \times 10^{-4}$ m, corresponding to mass transfer coefficient in the range $8.8 \times 10^{-6} < k < 2.3 \times 10^{-4}$ m s$^{-1}$, which is typical in RO modules. The CP calculated by the conventional film model is not dependent on the type of dominant salt. In contrast, the CP determined by the general Eq. 2 or Eq. 8 (the limiting case relevant to RO) varies widely for different dominant salts. These variations are due to the difference in diffusivities between the cations and anions constituting the dominant salts, as expressed in the parameter $\theta_\delta$. As $\theta_\delta$ deviates from 0 and approaches its boundaries (-1; 1), the CP of trace-ions further differs from the conventional film model. For example, using Eq. 2 or Eq. 8 (at $\delta = 100$ µm; $k \sim 9 \times 10^{-5}$ m s$^{-1}$), the trace $HPO_4^{2-}$ CP modulus in an $MgCl_2$ solution ($\theta_\delta = -0.39$) is 2.3x higher compared to the film model, while in an NaCl solution ($\theta_\delta = -0.21$) the CP modulus is only 1.4x higher (Fig. 2a). Similarly, trace $Ca^{2+}$ CP modulus in an $H_2SO_4$ solution ($\theta_\delta = 0.75$) is 2.5× higher compared to the film model (Fig. 2b). Illustratively, the 'slower' $SO_4^{2-}$ dominant ions pull the oppositely charged $Ca^{2+}$ trace-ions towards the membrane surface. In the three examples discussed above, $\theta_\delta$ has the same sign of the trace-ion charge, which results in increased CP. In contrast, when the trace-ion charge is opposite to $\theta_\delta$, its CP modulus is lower than the conventional film model. For example, the CP modulus of trace $Ca^{2+}$ in $MgCl_2$ solution ($\theta_\delta = -0.39$) is 2.5x lower, or illustratively, the 'faster' $Cl^-$ dominant ions pull the $Ca^{2+}$ trace-ions away from the membrane surface. The type of dominant salt can thus play a significant role in determining the extent of trace-ions CP.

As the boundary layer thickness increases (and the mass transfer coefficient decreases), CP aggravates, and deviations from the conventional film model become more prominent. A thick boundary layer may develop in practical membrane systems due to, e.g., low crossflow velocity in the last element of a membrane train; or in the first element after long-term accumulation of particulate fouling, inducing cake-enhanced CP [33,34]. For such severe CP conditions, where deviations from the film model are most prominent, the potential risks associated with high CP of trace ions – i.e., critical increase in solutes passage and chemical fouling – are most pronounced. Therefore, to assess these risks reliably, it is essential to account for the effect of electromigration on CP of trace-ions.

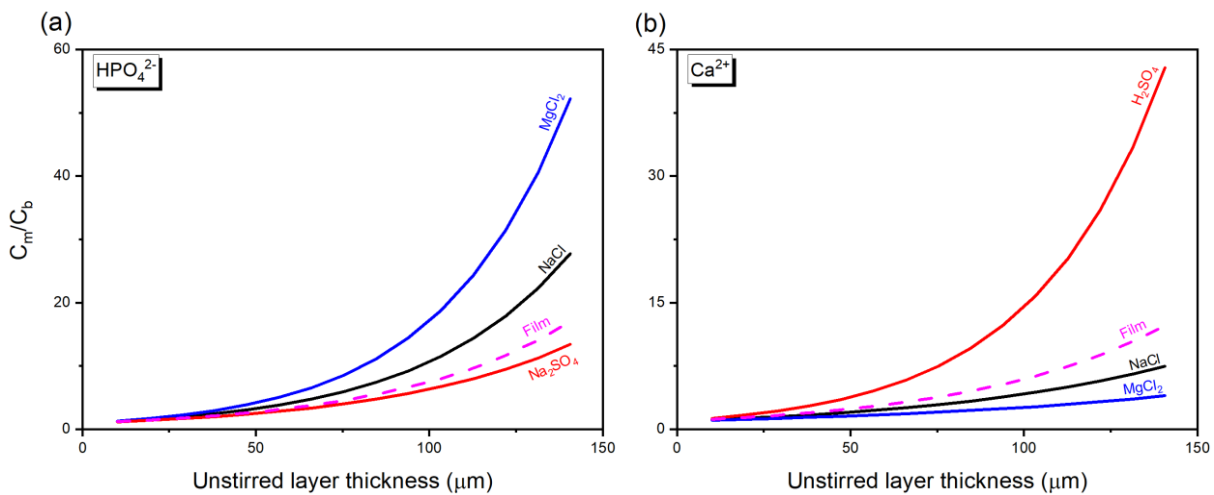

**Figure 2**. Concentration factor, $C_m/C_b$, on an RO membrane surface as a function of the unstirred layer thickness at $J_v$ = 50 LMH and a major salt concentration in the bulk solution of 85 mM for (a) 1.5 mM $HPO_4^{2-}$ as trace ion and $MgCl_2$, $NaCl$ or $Na_2SO_4$ as dominant salt, and (b) 5 mM $Ca^{2+}$ as trace ion in $H_2SO_4$, NaCl or MgCl$_2$ as dominant salt. Both conventional film-model (dashed line) and trace-ion theory for concentration polarization predictions are presented.

We further examined the effect of electromigration on CP in two case studies, representing environmentally relevant RO and NF membrane processes, as described above. In both cases (Fig. 3), variations in the CP modulus (Fig. 3a&c) calculated by either the general Eq. 2 or the conventional film model increased with the water flux. Consequently, variations in both the simulated precipitation potential (Fig. 3b) and passage (Fig. 3b&d) also increased with the water flux. For the RO case, the CP modulus calculated by the compact Eq. 8 closely matched the one calculated by Eq. 2, as the variations were lower than 4% for both the cations and anions simulated. In contrast, the CP modulus calculated by the conventional film theory (without electromigration) deviated by up to 80% for $Ca^{2+}$ and by 44% for $HPO_4^{2-}$.

The practical impact of accurate CP evaluation is evident in the RO case study (Fig. 3b). As seen in Fig. 3a, due to the negative value of $\theta_\delta$ for $NaCl$, the electromigrative flux act to increase the membrane surface concentration of $HPO_4^{2-}$ and decrease that of $Ca^{2+}$. Since these two divalent ions have the same absolute valence but opposite charge signs, Eq. 8 reveals that they decrease or increase by the same factor. Thereby, the ion activity product and the saturation index in the CP layer are not affected by electromigration, which is valid for any salt. However, as seen in Fig. 3b, the ACP precipitation potential – i.e., the amount of ACP to precipitate until thermodynamic equilibrium is reached – increases, indicating a higher risk of chemical fouling. The increase in precipitation potential compared to the film model (e.g., by 67% at $J_v$ = 14 µm/s) stems from the relative enrichment of $HPO_4^{2-}$ in the CP layer (e.g., by 72% at $J_v$ = 14 µm/s), which is at lower concentration and thus limit the precipitation reaction more than $Ca^{2+}$ does. Concerning decontamination efficiency, we show that the observed passage of $NH_4^+$ is overestimated by 26% when using the conventional film theory (Figure 3b) due to the overestimation of $NH_4^+$ membrane surface concentration (Fig. 3a inset). Notably, even for the relatively "open" RO membrane considered in case 1 ($NaCl$ rejection of 90-96%), Eq. 8, which assumes complete rejection, is in close agreement with the general Eq. 2, demonstrating the practical relevance of this simple extension to the film model. Thus, this case study suggests that accounting for electromigration in estimating CP for trace ions is essential. The compact Eq. 8 can be conveniently used for this purpose in dominant salt solutions.

The second case study included NF (with both a more open NF270 and a tight NF90 membrane) of arsenate-rich brackish groundwater (Table S1). The CP modulus determined by Eq. 9 (assuming complete rejection of trace-ions and partial rejection of the dominant salt) deviated from the one calculated by Eq. 2 by <8% for the NF90 membrane (Figure 3c), which is a significant improvement compared to the film theory in which CP modulus was 26% lower. However, for NF270 (Figure 3c), the CP modulus calculated by Eq. 9 was 22% higher than the one calculated by Eq. 2 (at 20 µm/s). This was expected for the more "open" NF270 membrane, having a moderate $HAsO_4^{2-}$ rejection (63%-73%) that considerably deviate from the assumption of complete ion rejection underlying Eq. 9. Accordingly, deviations (relatively to using Eq. 2) in the modeled $HAsO_4^{2-}$ passage appear when using both the film model (12%) and Eq. 9 (18%). Contrarily, in NF90, the $HAsO_4^{2-}$ passage modeled using Eq. 9 is in better agreement with the values predicted by Eq. 2 (deviation < 7.4%), while the film model deviates significantly (25.6%). Based on these two cases, we estimate that the compact Eq. 9 could be used to estimate CP of trace ions with rejection above 85%, while using the more general Eq. 2 is recommended for lower rejection.

### 4.1 Conclusion

Overall, the results imply that our simplified solutions could be a valuable addition to any model dealing with multicomponent mixtures in membrane separation. When reviewing the contemporary NF and RO literature, as briefly laid out in the introduction, it was evident that while some studies ignore the relevance of CP, many studies are aware of its importance – especially in multicomponent separation systems. However, while many studies mention the issue of CP in those systems, often a compromise is made to reduce the mathematical complexity involved in the proper account for electromigration. This oversimplification may lead to an incorrect, non-physical estimation of trace ions CP. The approximations presented here account for the electromigration factor and are easier to implement than any numerical solution proposed before. This is especially true for the compact limiting cases (Eq. 8 & 9), which naturally extend the film model and improves model reliability with a minimum effort, subject to assumptions often met in real applications.

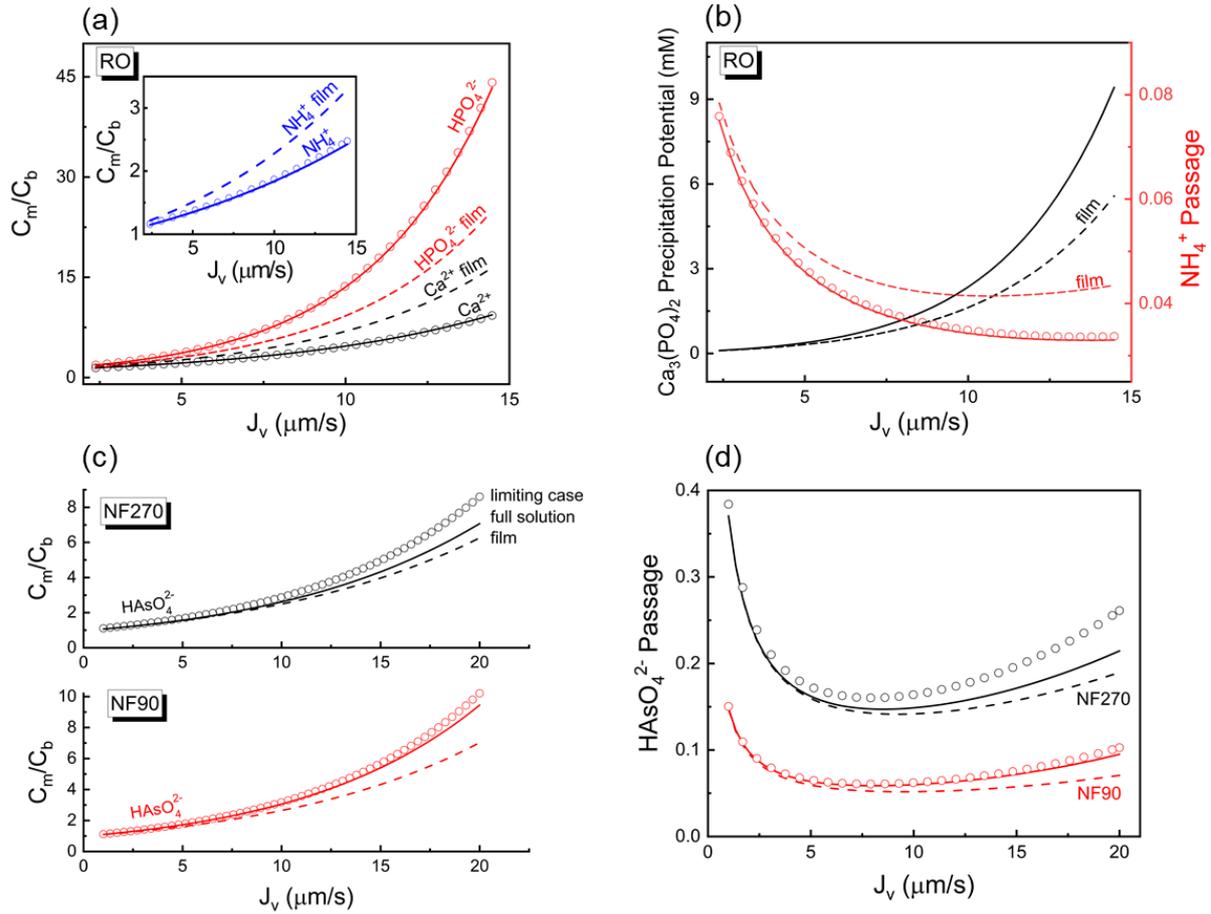

**Figure 3.** Each graph compares two calculations of concentration polarization: one including electromigration (solid lines) and one according to film-model (dashed lines). The numerical calculation of the limiting cases presented in this work is given by the circular symbols, where the first limiting case is used for case 1 and the second limiting case is used for case 2 with unstirred layer thickness, $\delta$, 150 µm and 64 µm, respectively. **Case 1**: (a) concentration polarization modulus of the trace ions $HPO_4^{2-}$, $Ca^{2+}$ and $NH_4^+$; (b) precipitation potential of $Ca_3(PO_4)_2$ and passage of trace $NH_4^+$. **Case 2**: (c) concentration polarization modulus of the trace ion $HAsO_4^{2-}$ in NF270 and NF90; (d) passage of trace $HAsO_4^{2-}$.

**Supplementary material**

A full derivation of the analytical solution for trace-ion membrane wall and permeate concentration is presented, as well as the classical film theory description. Feed concentrations for the discussed case studies are given, as well as the model parameters.

The Python codes for the presented case studies concerning RO ion passage and precipitation potential of ACP, as well as NF270 and NF90 ion passage are given.

**Acknowledgments**

This project received funding from the Israel Science Foundation (#2325/20). Y.O. would like to thank the Kreitman School of Advanced Graduate Studies for support through the Hightech, Biotech, and Chemotech scholarship program.

**References**

[1] V. Geraldes, M.D. Afonso, Prediction of the concentration polarization in the nanofiltration/reverse osmosis of dilute multi-ionic solutions, Journal of Membrane Science. 300 (2007) 20–27. https://doi.org/10.1016/j.memsci.2007.04.025.

[2] W.R. Bowen, A.W. Mohammad, Diafiltration by nanofiltration: Prediction and optimization, AIChE Journal. 44 (1998) 1799–1812. https://doi.org/10.1002/aic.690440811.

[3] Y.S. Oren, P.M. Biesheuvel, Theory of Ion and Water Transport in Reverse-Osmosis Membranes, Phys. Rev. Applied. 9 (2018) 024034. https://doi.org/10.1103/PhysRevApplied.9.024034.

[4] A.I. Cavaco Morão, A. Szymczyk, P. Fievet, A.M. Brites Alves, Modelling the separation by nanofiltration of a multi-ionic solution relevant to an industrial process, Journal of Membrane Science. 322 (2008) 320–330. https://doi.org/10.1016/j.memsci.2008.06.003.

[5] N. Fridman-Bishop, O. Nir, O. Lahav, V. Freger, Predicting the Rejection of Major Seawater Ions by Spiral-Wound Nanofiltration Membranes, Environ. Sci. Technol. 49 (2015) 8631–8638. https://doi.org/10.1021/acs.est.5b00336.

[6] J. López, O. Gibert, J.L. Cortina, Evaluation of an extreme acid-resistant sulphonamide based nanofiltration membrane for the valorisation of copper acidic effluents, Chemical Engineering Journal. 405 (2021) 127015. https://doi.org/10.1016/j.cej.2020.127015.

[7] H.C. van der Horst, J.M.K. Timmer, T. Robbertsen, J. Leenders, Use of nanofiltration for concentration and demineralization in the dairy industry: Model for mass transport, Journal of Membrane Science. 104 (1995) 205–218. https://doi.org/10.1016/0376-7388(95)00041-A.

[8] S. Bhattacharjee, J.C. Chen, M. Elimelech, Coupled model of concentration polarization and pore transport in crossflow nanofiltration, AIChE Journal. 47 (2001) 2733–2745. https://doi.org/10.1002/aic.690471213.

[9] S. Bhattacharjee, G.M. Johnston, A Model of Membrane Fouling by Salt Precipitation from Multicomponent Ionic Mixtures in Crossflow Nanofiltration, Environmental Engineering Science. 19 (2002) 399–412. https://doi.org/10.1089/109287502320963391.

[10] S. Déon, P. Dutournié, P. Fievet, L. Limousy, P. Bourseau, Concentration polarization phenomenon during the nanofiltration of multi-ionic solutions: Influence of the filtrated

solution and operating conditions, Water Research. 47 (2013) 2260–2272. https://doi.org/10.1016/j.watres.2013.01.044.

[11] A.R.D. Verliefde, E.R. Cornelissen, S.G.J. Heijman, J.Q.J.C. Verberk, G.L. Amy, B. Van der Bruggen, J.C. van Dijk, Construction and validation of a full-scale model for rejection of organic micropollutants by NF membranes, Journal of Membrane Science. 339 (2009) 10–20. https://doi.org/10.1016/j.memsci.2009.03.038.

[12] O. Coronell, B. Mi, B.J. Mariñas, D.G. Cahill, Modeling the Effect of Charge Density in the Active Layers of Reverse Osmosis and Nanofiltration Membranes on the Rejection of Arsenic(III) and Potassium Iodide, Environ. Sci. Technol. 47 (2013) 420–428. https://doi.org/10.1021/es302850p.

[13] S. Lee, R.M. Lueptow, Membrane Rejection of Nitrogen Compounds, Environ. Sci. Technol. 35 (2001) 3008–3018. https://doi.org/10.1021/es0018724.

[14] A. Azaïs, J. Mendret, E. Petit, S. Brosillon, Evidence of solute-solute interactions and cake enhanced concentration polarization during removal of pharmaceuticals from urban wastewater by nanofiltration, Water Research. 104 (2016) 156–167. https://doi.org/10.1016/j.watres.2016.08.014.

[15] K.M. Lim, N.F. Ghazali, Nanofiltration of binary palm oil/solvent mixtures: Experimental and modeling, Materials Today: Proceedings. (2020). https://doi.org/10.1016/j.matpr.2020.04.695.

[16] A. Giacobbo, E.V. Soares, A.M. Bernardes, M.J. Rosa, M.N. de Pinho, Atenolol removal by nanofiltration: a case-specific mass transfer correlation, Water Sci Technol. 81 (2020) 210–216. https://doi.org/10.2166/wst.2020.073.

[17] J. López, M. Reig, O. Gibert, J.L. Cortina, Integration of nanofiltration membranes in recovery options of rare earth elements from acidic mine waters, Journal of Cleaner Production. 210 (2019) 1249–1260. https://doi.org/10.1016/j.jclepro.2018.11.096.

[18] J. López, M. Reig, X. Vecino, O. Gibert, J.L. Cortina, From nanofiltration membrane permeances to design projections for the remediation and valorisation of acid mine waters, Science of The Total Environment. 738 (2020) 139780. https://doi.org/10.1016/j.scitotenv.2020.139780.

[19] Y.-A. Boussouga, H. Frey, A.I. Schäfer, Removal of arsenic(V) by nanofiltration: Impact of water salinity, pH and organic matter, Journal of Membrane Science. 618 (2021) 118631. https://doi.org/10.1016/j.memsci.2020.118631.

[20] J.R. Werber, A. Deshmukh, M. Elimelech, The Critical Need for Increased Selectivity, Not Increased Water Permeability, for Desalination Membranes, Environ. Sci. Technol. Lett. 3 (2016) 112–120. https://doi.org/10.1021/acs.estlett.6b00050.

[21] L. Pino, C. Vargas, A. Schwarz, R. Borquez, Influence of operating conditions on the removal of metals and sulfate from copper acid mine drainage by nanofiltration, Chemical Engineering Journal. 345 (2018) 114–125. https://doi.org/10.1016/j.cej.2018.03.070.

[22] A. Yaroshchuk, X. Martínez-Lladó, L. Llenas, M. Rovira, J. de Pablo, Solution-diffusion-film model for the description of pressure-driven trans-membrane transfer of electrolyte mixtures: One dominant salt and trace ions, Journal of Membrane Science. 368 (2011) 192–201. https://doi.org/10.1016/j.memsci.2010.11.037.

[23] M. Reig, E. Licon, O. Gibert, A. Yaroshchuk, J.L. Cortina, Rejection of ammonium and nitrate from sodium chloride solutions by nanofiltration: Effect of dominant-salt concentration on the trace-ion rejection, Chemical Engineering Journal. 303 (2016) 401–408. https://doi.org/10.1016/j.cej.2016.06.025.

[24] N. Pagès, M. Reig, O. Gibert, J.L. Cortina, Trace ions rejection tunning in NF by selecting solution composition: Ion permeances estimation, Chemical Engineering Journal. 308 (2017) 126–134. https://doi.org/10.1016/j.cej.2016.09.037.

[25] O. Nir, N.F. Bishop, O. Lahav, V. Freger, Modeling pH variation in reverse osmosis, Water Research. 87 (2015) 328–335. https://doi.org/10.1016/j.watres.2015.09.038.

[26] S. Phuntsho, S. Hong, M. Elimelech, H.K. Shon, Forward osmosis desalination of brackish groundwater: Meeting water quality requirements for fertigation by integrating nanofiltration, Journal of Membrane Science. 436 (2013) 1–15. https://doi.org/10.1016/j.memsci.2013.02.022.

[27] L. Birnhack, O. Nir, O. Lahav, Establishment of the Underlying Rationale and Description of a Cheap Nanofiltration-Based Method for Supplementing Desalinated Water with Magnesium Ions, Water. 6 (2014) 1172–1186. https://doi.org/10.3390/w6051172.

[28] S.C.N. Tang, L. Birnhack, Y. Cohen, O. Lahav, Selective separation of divalent ions from seawater using an integrated ion-exchange/nanofiltration approach, Chemical Engineering and Processing - Process Intensification. 126 (2018) 8–15. https://doi.org/10.1016/j.cep.2018.02.015.

[29] C.K. Diawara, L. Paugam, M. Pontié, J.P. Schlumpf, P. Jaouen, F. Quéméneur, Influence of Chloride, Nitrate, and Sulphate on the Removal of Fluoride Ions by Using Nanofiltration Membranes, Separation Science and Technology. 40 (2005) 3339–3347. https://doi.org/10.1080/01496390500423706.

[30] O. Labban, C. Liu, T.H. Chong, J.H. Lienhard V, Fundamentals of low-pressure nanofiltration: Membrane characterization, modeling, and understanding the multi-ionic interactions in water softening, Journal of Membrane Science. 521 (2017) 18–32. https://doi.org/10.1016/j.memsci.2016.08.062.

[31] Eric Jones, Travis Oliphant, Pearu Peterson and others, SciPy: Open Source Scientific Tools for Python, 2001. http://www.scipy.org.

[32] M.J.J.M. van Kemenade, P.L. de Bruyn, A kinetic study of precipitation from supersaturated calcium phosphate solutions, Journal of Colloid and Interface Science. 118 (1987) 564–585. https://doi.org/10.1016/0021-9797(87)90490-5.

[33] R. Bian, K. Yamamoto, Y. Watanabe, The effect of shear rate on controlling the concentration polarization and membrane fouling, Desalination. 131 (2000) 225–236. https://doi.org/10.1016/S0011-9164(00)90021-3.

[34] E.M.V. Hoek, M. Elimelech, Cake-Enhanced Concentration Polarization: A New Fouling Mechanism for Salt-Rejecting Membranes, Environ. Sci. Technol. 37 (2003) 5581–5588. https://doi.org/10.1021/es0262636.

Graphical abstract

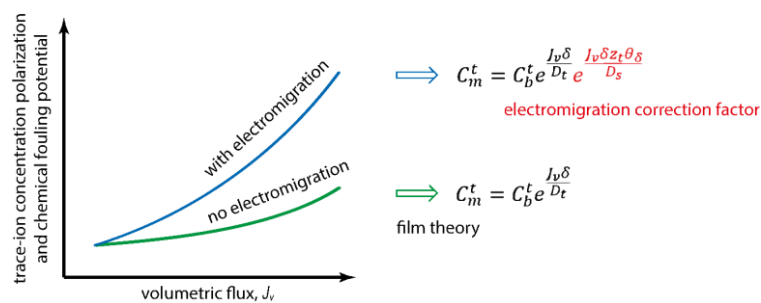